\documentclass[preprint, amsmath,amssymb,aps,natbib,prf]{revtex4-2}
\usepackage{geometry}
\usepackage{comment}
\usepackage{caption}
\usepackage{enumerate}
\usepackage{amsmath}
\usepackage{xcolor}
\usepackage{mathrsfs}
\usepackage{tabularx}
\usepackage{graphicx}
\usepackage{dcolumn}
\usepackage{appendix}
\usepackage{natbib}
\usepackage{bm}

\usepackage[mathlines]{lineno}
\usepackage{xcolor}
\usepackage{color}
\usepackage[colorlinks=true,linktoc=page,citecolor=blue,linkcolor=blue,]{hyperref}
\usepackage{graphicx}
\usepackage{subcaption}
\usepackage[section]{placeins}
\usepackage[export]{adjustbox}
\usepackage{float}

\begin{document}

\title{Stationary and non-stationary energy cascades in homogeneous ferrofluid turbulence 
		} 
\author{Sukhdev Mouraya}
\homepage{Author to whom correspondence should be addressed: sukhdev@iitk.ac.in}
\author{Nandita Pan}
\author{Supratik Banerjee}
\affiliation{
 Department of Physics, Indian Institute of Technology Kanpur, Uttar Pradesh, 208016 India
}%

\date{\today}
\begin{abstract}

The nonlinear transfer rate of the total energy (transfer rate of kinetic energy + transfer rate due to the work done by the magnetization) for an incompressible turbulent ferrofluid system is studied under the assumption of statistical homogeneity. Using the formalism of the two-point correlators, an exact relation connecting the second-order statistical moments to the average energy injection rate is derived for the scale-to-scale transfer of the total energy. We validate the universality of the exact relation through direct numerical simulations for stationary and non-stationary cascade regimes. For a weak external magnetic field, both kinetic and the total energy cascade with nearly the same cascade rate. A stationary cascade regime is achieved and hence a good agreement between the exact energy transfer rate and the average energy injection is found. Due to the rapid alignment of the ferrofluid particles in the presence of strong external fields, the turbulence dynamics becomes non-stationary. Interestingly, there too, both kinetic and the total energy exhibit inertial range cascades but with different cascade rates which can be explained using the non-stationary form of our derived exact relation. 
\end{abstract}

\maketitle
\vspace{-1.0cm}
{\centering \section{\label{sec:level1}Introduction}}

\vspace{-0.25cm}
To date, turbulence is one of the most challenging problems of physics that needs to be completely understood. In common fluids, turbulence arises from the nonlinear interactions among neighbouring fluid layers, which leads to the formation of eddy-like structures of various sizes thus conceiving a wide range of length scales. Energy is fed into the system at large scales and is dissipated at very small scales. Well inside the intermediate inertial length scales, free from the effects of forcing and dissipation, a fully developed turbulence is characterized by a universal cascade of kinetic energy across the scales with a constant transfer rate $\varepsilon$. For homogeneous and isotropic turbulence, $\varepsilon$ can be exactly expressed in terms of the third-order moments of two-point velocity fluctuations. For a turbulent magnetohydrodynamic fluid, the total energy (sum of kinetic and magnetic energy) cascades throughout the inertial range. The respective energy transfer rate is expressed in terms of two-point fluctuations of both velocity and magnetic field. 

Derivation of the exact relations connecting $\varepsilon$ and the two-point fluctuations was pioneered by Kolmogorov \citep{Kolmogorov1941} for a homogeneous and isotropic incompressible hydrodynamic fluid. Following Kolmogorov, similar exact laws were also derived for different fluid systems including magnetohydrodynamic (MHD) turbulence \cite{Karman1938, Politano1998, politano1998dynamical, Politano2003, Galtier2008}. Without the explicit assumption of isotropy, differential exact relations involving the divergence of the third-order moments have been derived for several other systems including the ones mentioned above \citep{Monin1975, Antonia1997, Podesta2008, Augier2012, Nandita2022}. This formalism is also extended to compressible systems, where the $\varepsilon$ can be written as a divergence of the third-order moments plus some source terms \cite{Banerjee2011, Banerjee2013, Banerjee2014, Banerjee2017}. An alternative form of the exact relation involving the second-order statistical moments is recently proposed for incompressible HD and MHD turbulence \cite{Banerjee2016, Banerjee2016b, Nahuel2019, Banerjee2018}. This form is particularly interesting to cases where $\varepsilon$ can not be written purely in terms of the divergence of the two-point fluctuations (e.g., energy cascade in incompressible Hall-MHD turbulence, turbulent compressible flows, etc.) \cite{Banerjee2018, Banerjee2020}. 
One can then be interested to know if a similar type of universal cascade can also be found in more complex fluids such as ferrofluids. Ferrofluids are complex synthetic liquids that contain nano-sized ferromagnetic particles suspended in a carrier liquid (water, oil, and other organic solvents, etc.). Ferrofluid particles interact via attractive van der Waals forces and dipole-dipole interactions, which may result in aggregations. To prevent immediate aggregation, these ferro-particles are, in general, coated with a surfactant \citep{Shliomis1972, Zebib1996, Rosensweig1997, Kaloni2004}. In the absence of an external magnetic field, the particles are randomly orientated and the fluid has no net magnetism. However, in the presence of an external field ($\textbf{H}_0$), the magnetic particles respond by forming chain-like structures along the field lines  \citep{Huang2005, Buschmann2020}. It is often necessary to slow down such agglomeration of the particles and to break the chain-like structure to obtain a sustained ferrofluid emulsion. This can be achieved by making the system turbulent.

Chain-like structures and alignments in the presence of an external field resist the mobility of a ferrofluid. A ferrofluid therefore exhibits less turbulence than ordinary HD or MHD fluid \citep{Kaloni2004, Shliomis1994, Bacri1995, BAFinlayson2013}. However, when confronted with rapid flow velocities at a very large Reynolds number ferrofluid can still exhibit varying degrees of turbulent behavior. In such cases, the ferrofluid's response to turbulent flows can be significantly influenced by the concentration of magnetic particles and the external magnetic field. In particular, a stronger external magnetic field would tend to quickly align the ferro-particles thereby competing with the development of turbulence in the flow \citep{Adrian1981, Schumacher2003, Schumacher2008, Schumacher2010, Schumacher2011, Altmeyer2015}. Explorating turbulence within ferrofluid is imperative for investigating the implications of magnetic fields on heat transfer phenomena, encompassing flow management, augmenting heat transfer processes, and mechanisms dedicated to noise reduction through convection \citep{Raj1995, Snyder2003, LiliSha2017, Kole2021}. 

Although a number of theoretical and experimental studies have been carried out on ferrofluid flows \cite{BAFinlayson1970, Bacri1995, Felderhof1999, Kaloni2004, BAFinlayson2013}, only a few of them are dedicated to study the turbulent properties of such a flow \citep{Adrian1981, Schumacher2008, Schumacher2010, Schumacher2011, Altmeyer2015}. In particular, for homogeneous turbulence, those studies investigated the evolution of the turbulent kinetic energy (both translational and rotational) and also the corresponding power spectra \cite{Schumacher2003, Schumacher2008, Schumacher2010}. In a turbulent ferrofluid, the sum of kinetic energy and work done due to magnetization is an inviscid invariant and is therefore expected to exhibit a universal cascade with a constant flux rate ($\varepsilon$) inside the inertial range. However, the possibility of a universal cascade in terms of the statistical moment of field fluctuations has not been studied until recently where an exact relation for homogeneous ferrofluid turbulence has been derived \cite{Mouraya2019}. Unlike incompressible HD, MHD and binary fluid dynamics, $\varepsilon$ in ferrofluid turbulence cannot be expressed as a divergence of two-point increments and instead, an alternative form similar to \cite{Banerjee2016, Banerjee2016b} of the exact relation has been obtained for the energy cascade in ferrofluid turbulence.

In the current study, we revisit our previous work \cite{Mouraya2019} with a realistic assumption of negligible particle size and re-derive a reduced form of the exact relation for the total energy transfer.
The derived law is then numerically tested with 3D direct numerical simulations (DNS) ranging from $128^3$ to $512^3$ grid points to find the signature of scale-independent energy transfer rate. Finally, stationary and non-stationary cascades are studied by varying the strength of $\textbf{H}_0$ and the effect of an external field on the nature of turbulent cascade in the ferrofluids is discussed. 

The paper is organized as follows. In Sec~\ref{sec:level2} the governing equations and conservation of total energy for the ferrofluid system are described whereas Sec.~\ref{sec:level3} contains the derivation of the exact relation for negligible ferrofluid particle size. In Sec.~\ref{sec:level5}, we present the numerical methods and other simulation details. In Sec.~\ref{sec:level}, we present and discuss our findings. Finally, in Sec. \ref{sec:level6}, we summarize and conclude.

{\centering \section{\label{sec:level2}Model and Inviscid Invariant}} 
{\centering \subsection{Basic equations}}

The governing equations for incompressible ferrofluid (the constant density is normalized to unity) consist of the evolution equations for linear momentum, angular momentum and magnetization \citep{Schumacher2008, Mouraya2019}:
\begin{align}
 \left( \partial_t + \bm{v} \cdot \boldsymbol{\nabla} \right) \bm{v}  &=- \boldsymbol{\nabla} p  + \nu \nabla^2 \bm{v} + \mu_0 (\textbf{M} \cdot \boldsymbol{\nabla}) \textbf{H}- \zeta \boldsymbol{\nabla} \times (  \boldsymbol{\Omega}-2\boldsymbol{\omega}) 
, \label{1} \\
 I \left( \partial_t + \bm{v} \cdot\boldsymbol{\nabla} \right) \boldsymbol{\omega}  &=  \mu_0 {(\textbf{M} \times \textbf{H})} +\eta \nabla^2 \boldsymbol{\omega} + 2 \zeta (\boldsymbol{\Omega}-2\boldsymbol{\omega}), \label{2}	\\
(\partial_t + \bm{v} \cdot \boldsymbol{\nabla}) \textbf{M} &=\boldsymbol{\omega} \times \textbf{M}-\frac{1}{\tau}(\textbf{M}-\textbf{M}_{eq}), \label{3} \\
\boldsymbol{\nabla} \cdot \bm{v} &= 0, \hspace{0.2cm} \boldsymbol{\nabla} \times \textbf{H} = \textbf{0}, \label{4}
\end{align}
where $\bm{v}$ is the velocity of the ferrofluid, $\boldsymbol{\Omega} = \boldsymbol{\nabla} \times \bm{v}$ is the vorticity, $p$ is the fluid pressure, $\boldsymbol\omega$ is the ferrofluid particle spin rate, $\textbf{M}$ is the magnetization vector, $\textbf{H}$ is the magnetic field vector, $I$ is the  moment of inertia per unit mass for a ferrofluid particle, $\nu$ is the kinematic viscosity, $\zeta$ is the vortex viscosity,
$\eta$ is the spin viscosity and $\tau$ is the relaxation time.
The equilibrium magnetization is given by $\textbf{M}_{eq} = M_s L(\xi) {\textbf{H}}/{H}$, where 
$ L(\xi) = (\xi \coth(\xi)-{1})/{\xi} $ is the Langevin function with $\xi = {\mu_0 m H}/{k_B T} $ for a ferrofluid at temperature $T$. The parameters $M_s$ and $m$ are the magnitudes of saturation magnetization and magnetic moment of a single ferrofluid particle respectively. For small values of $\xi$, one can write $\textbf{M}_{eq} = \chi \textbf{H}$, where $\chi$ is magnetic susceptibility. Due to incompressibility, $\bm{v}$ is divergence-less and in the absence of any free current $\textbf{H}$ is irrotational. Since, the magnetic flux (\textbf{B}) is divergence-free, the evolution of the magnetic field can directly be obtained from that of the magnetization as $\textbf{B} = \mu_0 (\textbf{H} + \textbf{M})$ and hence one can write $\boldsymbol{\nabla} \cdot \textbf{M} = - \boldsymbol{\nabla} \cdot \textbf{H}$.

Unlike our previous work \cite{Mouraya2019}, here we incorporate the fact that the specific moment of inertia $ I (\sim 10^{-16} m^2)$ and the spin viscosity $\eta~(\sim 10^{-15} kg \ m \ s^{-1})$ are very small and can practically be neglected with respect to the other terms in the evolution equation of angular momentum \citep{Schumacher2008}. Hence, the angular momentum Eq.~\eqref{2} reduces to
\begin{equation}
    \boldsymbol{\omega} = \frac{\boldsymbol{\Omega}}{2} +\frac{\mu_0}{4 \zeta} (\textbf{M} \times \textbf{H}). \label{5}
\end{equation}
Using Eq. \eqref{5} in Eqs. \eqref{1} and \eqref{3}, one obtains
\begin{align}
 ( \partial_t + \bm{v} \cdot \boldsymbol{\nabla}) \bm{v} &=- \boldsymbol{\nabla} p + \nu \nabla^2 \bm{v} + \mu_0 (\textbf{M} \cdot \boldsymbol{\nabla}) \textbf{H} + \frac{\mu_0}{2} \boldsymbol{\nabla} \times (\textbf{M} \times \textbf{H})
, \label{6}\\	
(\partial_t + \bm{v} \cdot \boldsymbol{\nabla}) \textbf{M} &= \frac{1}{2} (\boldsymbol{\Omega} \times \textbf{M}) + \frac{\mu_0}{4 \zeta} (\textbf{M} \times \textbf{H}) \times \textbf{M}-\frac{1}{\tau}(\textbf{M}-\chi \textbf{H}), \label{7}
\end{align} 
which constitute the revised set of governing equations for the ferrofluid turbulence.

\vspace{1.0cm}
\begin{center}
{\subsection{Conservation of total energy  and Nondimensionalization}}
\end{center}
\vspace{-1.25cm}
Similar to ordinary fluids, ferrofluid equations also conserve the total energy in the absence of viscous terms.
The total energy consists of the kinetic energy of the fluid and the internal energy resulting from the work performed by the ferrofluid particles in response to the external magnetic field. Using  Eq.~\eqref{6}, the kinetic energy evolution equation can be written as
\begin{align}
   &\partial_t E_{kin}  =  \partial_t \left( \frac{ v^2}{2}\right) \nonumber \\ &= -\boldsymbol{\nabla} \cdot \left[ \left( \frac{v^2}{2} + p \right) \bm{v} - \frac{\mu_0}{2}  (\textbf{M} \times \textbf{H}) \times \bm{v} \right] + \frac{\mu_0}{2}\left[(\textbf{M} \times \textbf{H}) \cdot \boldsymbol{\Omega} 
   + 2 \bm{v} \cdot (\textbf{M} \cdot {\boldsymbol{\nabla}}) \textbf{H}\right] - \nu \Omega^2 
   .
\end{align}
Again using Eq.~\eqref{7}, the differential work performed by the ferrofluid particles due to $\textbf{H}$ is given by $d E_{mag} = - \textbf{H} \cdot d\textbf{M}$ and hence the corresponding evolution equation is given by
\begin{align}	
&\partial_t E_{mag} = -\textbf{H} \cdot \partial_t \textbf{M}\nonumber \\& =  \textbf{H} \cdot ((\bm{v} \cdot \boldsymbol{\nabla}) \textbf{M}) - \frac{1}{2} \textbf{H} \cdot(\boldsymbol{\Omega} \times \textbf{M})  - \frac{\mu_0}{4\zeta} \textbf{H} \cdot ((\textbf{M} \times \textbf{H}) \times \textbf{M}) +\frac{1}{\tau} \textbf{H} \cdot (\textbf{M}-\chi \textbf{H}) \nonumber \\ 
& =  \textbf{H} \cdot ((\bm{v} \cdot \boldsymbol{\nabla}) \textbf{M}) - \frac{1}{2} \textbf{H} \cdot(\boldsymbol{\Omega} \times \textbf{M}) - \frac{\mu_0}{4 \zeta} (\textbf{M} \times \textbf{H})^2 + \frac{1}{\tau} \textbf{H} \cdot (\textbf{M}-\chi \textbf{H})\nonumber \\ 
& =  \textbf{H} \cdot ((\bm{v} \cdot \boldsymbol{\nabla}) \textbf{M}) - \frac{1}{2} \textbf{H} \cdot(\boldsymbol{\Omega} \times \textbf{M}) - \frac{\zeta}{\mu_0} ( 2 \boldsymbol{\omega} - \boldsymbol{\Omega} )^2 +\frac{1}{\tau} \textbf{H} \cdot (\textbf{M}-\chi \textbf{H}).
\end{align}

The evolution equation of the total energy is given by
\begin{equation}
 \int \partial_t \left(E_{kin} + E_{mag}\right) d \tau  = - \int \boldsymbol{\nabla} \cdot \left[ \left( \frac{v^2}{2} + p + \mu_0 \textbf{M} \cdot \textbf{H} \right) \bm{v} - \frac{\mu_0}{2} (\textbf{M} \times \textbf{H}) \times \bm{v} \right] d \tau + d 
, \label{10}
\end{equation}
where dissipation term is:
\begin{align}
    d &= \int \left( -\nu \Omega^2 + \frac{\mu_0}{\tau} \textbf{H} \cdot (\textbf{M} - \chi \textbf{H}) - \zeta (  \boldsymbol{\Omega} -2 \boldsymbol{\omega} )^2 \right) d \tau.
\end{align}
The first term in the RHS of Eq.~\eqref{10} vanishes by the use of the Gauss divergence theorem and finally ignoring the dissipation terms, we obtain the conservation of the total energy. In viscous flows, where the dissipative effects are not negligible, the left-hand side of Eq.~\eqref{10} does not vanish. In such a situation, one needs to add an additional injection term $f$ which would then lead to the conservation of energy by balancing the dissipation as $d= -f$. In the next section, we shall see if a constant flux of energy can be obtained by assuring the driving only at large scales and dissipation only at small scales. 

{\centering\section{\label{sec:level3}Derivation of Exact relations}}

In this section, using two-point statistics, we derive an exact relation for the transfer of total energy within the inertial range.
For the sake of numerical implementation, it is necessary to cast the constitutive relation \eqref{5} and the governing Eqs.~\eqref{6} and \eqref{7} in terms of dimensionless starred variables as
\begin{align}
\boldsymbol{\omega}^* &= \frac{1}{2}\boldsymbol{\Omega}^* +\frac{Re}{2.2} (\textbf{M}^* \times \textbf{H}^*)\label{24},\\
\partial_{t^*} \bm{v}^* &= \bm{v}^* \times \boldsymbol{\Omega}^* - \boldsymbol{\nabla}^* \left(p^* + \frac{v^{*2}}{2}\right) + \frac{1}{Re} \nabla^{*2} \bm{v}^* + (\textbf{M}^* \cdot \boldsymbol{\nabla}^*) \textbf{H}^* + \frac{1}{2} \boldsymbol{\nabla}^* \times (\textbf{M}^* \times \textbf{H}^*) 
, \label{22a} \\ 
\partial_{t^*} \textbf{M}^* &= - (\bm{v}^* \cdot \boldsymbol{\nabla}^*) \textbf{M}^* + \frac{1}{2} (\boldsymbol{\Omega}^* \times \textbf{M}^*) + \frac{Re}{2.2} (\textbf{M}^* \times \textbf{H}^*) \times \textbf{M}^*-\frac{1}{\Gamma}(\textbf{M}^*-\chi \textbf{H}^*), \label{22b} 
\end{align}
where we used $\zeta = 0.55\nu$ \citep{Schumacher2008}, $ \bm{v} = v_{rms} \bm{v}^*$, $\textbf{M} = \sqrt{\frac{1}{\mu_0}} v_{rms} \textbf{M}^*$, $\textbf{H} = \sqrt{\frac{1}{\mu_0}} v_{rms} \textbf{H}^*$, $t = \frac{l_0}{v_{rms}} t^*$, $p =  v_{rms}^2 p^*$, $\Gamma = \frac{v_{rms}}{l_0} \tau $, $Re = \frac{l_0 v_{rms}}{\nu}$ is the large-scale Reynolds number, with $l_0$ representing the box size and $v_{rms}$ the root mean square velocity. In order to simplify the notations, we shall omit the stars from the dimensionless variables hereinafter. To achieve a sustained turbulent flow, the system is driven by a large-scale forcing $\bm{f}_{v}$ (delta correlated in time) in the momentum equation.
The evolution equation of the energy correlation function is then given by (using Eqs.~\eqref{22a} and ~\eqref{22b})
\begin{align}
\partial_t \mathcal{R} &= \frac{1}{2}  \left< (\bm{v}' \cdot \partial_t \bm{v} +  \bm{v} \cdot \partial_t \bm{v}')  - \left( \textbf{H} \cdot \partial_t \textbf{M}' + \textbf{H}' \cdot \partial_t \textbf{M} \right) \right> \nonumber \\ 
&= \frac{1}{2}\left< \bm{v}' \cdot (\bm{v} \times \boldsymbol{\Omega}) + \bm{v}  \cdot (\bm{v}' \times \boldsymbol{\Omega}') - \bm{v}  \cdot \boldsymbol{\nabla}' \left(p' + \frac{v^{'2}}{2} \right) - \bm{v}'  \cdot \boldsymbol{\nabla} \left( p + \frac{v^{2}}{2} \right) - \frac{1}{2} \textbf{H}' \cdot (\boldsymbol{\Omega} \times \textbf{M} ) \right. \nonumber \\ 
& - \left. \frac{1}{2} \textbf{H} \cdot (\boldsymbol{\Omega}' \times \textbf{M}') + \bm{v} \cdot (\textbf{M}' \cdot \boldsymbol{\nabla}') \textbf{H}' + \bm{v}' \cdot (\textbf{M} \cdot \boldsymbol{\nabla}) \textbf{H} +\textbf{H}' \cdot(\bm{v} \cdot \boldsymbol{\nabla}) \textbf{M}+  \textbf{H} \cdot(\bm{v}' \cdot \boldsymbol{\nabla}') \textbf{M}'\right.
\nonumber \\ 
& + \left. \frac{1}{2} \bm{v}' \cdot (\boldsymbol{\nabla} \times (\textbf{M} \times \textbf{H}))  + \frac{1}{2} \bm{v} \cdot (\boldsymbol{\nabla}' \times (\textbf{M}' \times \textbf{H}'))\right> + \left<\frac{1}{Re} \bm{v} \cdot \nabla'^2 \bm{v}' + \frac{1}{Re} \bm{v}' \cdot \nabla^2 \bm{v} \right.
\nonumber \\ 
& - \left. \frac{Re}{2.2} \textbf{H}' \cdot ((\textbf{M} \times \textbf{H}) \times \textbf{M}) - \frac{Re}{2.2} \textbf{H} \cdot ((\textbf{M}' \times \textbf{H}') \times \textbf{M}') +  \frac{1}{\Gamma} \textbf{H}' \cdot (\textbf{M} - \chi \textbf{H}) \right.
\nonumber \\ 
& + \left. \frac{1}{\Gamma} \textbf{H} \cdot (\textbf{M}' - \chi \textbf{H}')
+ \bm{v}' \cdot \bm{f}_v + \bm{v} \cdot \bm{f}'_v \right> \label{corr}
\end{align}
where unprimed and primed quantities represent the variables at the point $\bm{x}$ and $\bm{x}'(\equiv \bm{x} + \boldsymbol{\ell})$ respectively, $\boldsymbol{\ell}$ is the increment vector and $\langle\cdot\rangle$ denotes the ensemble average, which is equivalent to the space average due to statistical homogeneity. 
Again, using incompressibility and homogeneity, one can show $\left< \bm{v}' \cdot \boldsymbol{\nabla} \left( \frac{v^2}{2} + p \right) \right> = \left< \bm{v} \cdot \boldsymbol{\nabla}' \left( \frac{v'^2}{2} + p' \right) \right> = 0$. 
Eq.~\eqref{corr} can further be simplified as 
\begin{align}
\partial_t \mathcal{R} &= \frac{1}{2} \left<- \delta(\bm{v} \times \boldsymbol{\Omega} ) \cdot \delta\bm{v}  - \delta\bm{v} \cdot \delta ((\textbf{M} \cdot \boldsymbol{\nabla}) \textbf{H}) - \delta \textbf{H} \cdot \delta ((\bm{v} \cdot \boldsymbol{\nabla}) \textbf{M}) - \frac{1}{2} \delta\bm{v} \cdot \delta (\boldsymbol{\nabla} \times (\textbf{M} \times \textbf{H})) \right. \nonumber \\ 
& + \left. \frac{1}{2} \delta \textbf{H} \cdot \delta(\boldsymbol{\Omega} \times \textbf{M})\right> + \left<\frac{1}{Re} \bm{v} \cdot \nabla'^2 \bm{v}' + \frac{1}{Re} \bm{v}' \cdot \nabla^2 \bm{v} - \frac{Re}{2.2} \textbf{H}' \cdot ((\textbf{M} \times \textbf{H}) \times \textbf{M}) \right.
\nonumber \\ 
&  \left.- \frac{Re}{2.2} \textbf{H} \cdot ((\textbf{M}' \times \textbf{H}') \times \textbf{M}') +  \frac{1}{\Gamma} \textbf{H}' \cdot (\textbf{M} - \chi \textbf{H}) + \frac{1}{\Gamma} \textbf{H} \cdot (\textbf{M}' - \chi \textbf{H}')
+ \bm{v}' \cdot \bm{f}_v + \bm{v} \cdot \bm{f}'_v \right>, \label{Corr1}
\end{align}
where we have used the following relations (obtained under the assumption of statistical homogeneity) 
\begin{align}
    & (i) \left<(\bm{v} \times \boldsymbol\Omega ) \cdot \bm{v}'+(\bm{v}' \times \boldsymbol\Omega') \cdot \bm{v} \right> =- \left< \delta(\bm{v} \times \boldsymbol\Omega)\cdot \delta \bm{v} \right> \\
    & (ii) \left< \bm{v} \cdot (\textbf{M}' \cdot \boldsymbol{\nabla}') \textbf{H}' + \bm{v}'\cdot (\textbf{M} \cdot \boldsymbol{\nabla}) \textbf{H} + \textbf{H} \cdot (\bm{v}' \cdot \boldsymbol{\nabla}') \textbf{M}' + \textbf{H}' \cdot (\bm{v} \cdot \boldsymbol{\nabla}) \textbf{M} \right> \nonumber \\ 
    \,& = \left< - \delta \bm{v} \cdot \delta((\textbf{M} \cdot \boldsymbol{\nabla}) \textbf{H}) - \delta \textbf{H} \cdot \delta((\bm{v} \cdot \boldsymbol{\nabla}) \textbf{M}) \right> \\ 
     & (iii)  \left< \textbf{H}' \cdot (\boldsymbol{\Omega} \times \textbf{M}) + \textbf{H} \cdot (\boldsymbol{\Omega}' \times \textbf{M}') - \bm{v}' \cdot (\boldsymbol{\nabla} \times (\textbf{M} \times \textbf{H})) - \bm{v} \cdot (\boldsymbol{\nabla}' \times (\textbf{M}' \times \textbf{H}')) \right> \nonumber \\
     & = \left< \delta \bm{\Omega} \cdot \delta ( (\textbf{M} \times \textbf{H})) - \delta \textbf{H} \cdot \delta (\boldsymbol{\Omega} \times \textbf{M}) \right>.
\end{align}
To see the effect of the applied external field $\textbf{H}_0$\ on the energy transfer, we apply a uniform external field of strength $H_0$ along $\hat{z}$ direction. Decomposing the total magnetic field as $\textbf{H} = \textbf{H}_0 + \textbf{\~H} = H_0 \hat{z} + \textbf{\~H}$, we obtain
\begin{align}
\partial_t \mathcal{R} &= \frac{1}{2} \left<- \delta(\bm{v} \times \boldsymbol{\Omega} ) \cdot \delta\bm{v}  - \delta\bm{v} \cdot \delta ((\textbf{M} \cdot \boldsymbol{\nabla}) \textbf{\~H}) - \delta \textbf{\~H} \cdot \delta((\bm{v} \cdot \boldsymbol{\nabla}) \textbf{M}) - \frac{1}{2} \delta\bm{\Omega} \cdot \delta (\textbf{M} \times \textbf{\~H}) \right. \nonumber \\ 
& + \left. \frac{1}{2} \delta \textbf{\~H} \cdot \delta(\boldsymbol{\Omega} \times \textbf{M})\right> + \left<\frac{1}{2} \textbf{H}_0 \cdot (\delta \bm{\Omega} \times \delta\textbf{M} )\right> + D + F,\label{Corr3}
\end{align}
where $D$ consists of the two-point dissipative terms
\begin{align}
    D &= \frac{1}{2}\left< \frac{1}{Re} \bm{v} \cdot \nabla'^2 \bm{v}' + \frac{1}{Re} \bm{v}' \cdot \nabla^2 \bm{v} - \frac{Re}{2.2} \textbf{\~H}' \cdot ((\textbf{M} \times \textbf{H}) \times \textbf{M}) - \frac{Re}{2.2} \textbf{\~H} \cdot ((\textbf{M}' \times \textbf{H}') \times \textbf{M}') \right. \nonumber \\ & + \left. \frac{1}{\Gamma} \textbf{\~H}' \cdot (\textbf{M} - \chi \textbf{H}) + \frac{1}{\Gamma} \textbf{\~H} \cdot (\textbf{M}' - \chi \textbf{H}') \right> \label{28} 
\end{align}
and $F$ consists of purely the large-scale contributions, where
\begin{align}
    F &= \frac{1}{2} \left< \bm{v}' \cdot \bm{f}_v + \bm{v} \cdot \bm{f}'_v \right \rangle - \left\langle\frac{Re}{2.2} \textbf{H}_0 \cdot ((\textbf{M} \times \textbf{H}) \times \textbf{M}) -  \frac{1}{\Gamma} \textbf{H}_0 \cdot (\textbf{M} - \chi \textbf{H}) \right>
    \label{29}.
\end{align}
Assuming a statistically stationary state where $\partial_t \mathcal{R} =0$ and ignoring the dissipative effects inside the inertial range, finally, we obtain the exact relation as
\begin{align} 
 A (\bm{\ell}) = A_1 (\bm{\ell}) + A_2 (\bm{\ell}) &= 2 \varepsilon, \label{23}
\end{align}
where
{\fontsize{10.5pt}{10.5pt}\selectfont
\begin{align}
    A_1 (\bm{\ell}) &= \left< \delta \bm{v} \cdot \left[ \delta(\bm{v} \times \boldsymbol\Omega)  +  \delta((\textbf{M} \cdot \boldsymbol{\nabla}) \textbf{\~H}) \right] + \delta \textbf{\~H} \cdot \left[ \delta ((\bm{v} \cdot \boldsymbol{\nabla}) \textbf{M}) -   \frac{\delta (\boldsymbol{\Omega} \times \textbf{M})}{2} \right] + \delta \boldsymbol{\Omega} \cdot  \frac{\delta (\textbf{M} \times \textbf{\~H})}{2} \right> \\
 A_2 (\bm{\ell}) &= - \left<  \frac{1}{2}  \textbf{H}_0\cdot (\delta \boldsymbol{\Omega} \times \delta \textbf{M}) \right>, \label{34}
\end{align}}

and $\varepsilon = F \approx \left< \bm{v} \cdot \bm{f}_v - \textbf{H}_0 \cdot \left[ \frac{Re}{2.2} ((\textbf{M} \times \textbf{H}) \times \textbf{M}) - \frac{1}{\Gamma} (\textbf{M} - \chi \textbf{H}) \right]\right> = \varepsilon_{inj} - \varepsilon_{H_0} $  is the mean energy injection rate.

Eq.~\eqref{23} is the main analytical result of this paper. It gives the energy cascade rate for ferrofluid turbulence, where the ferrofluid particles are of negligible size. As is evident from Eq.~\eqref{23}, the exact relation is not completely free from the mean-field effect. As shown above, the external magnetic field ($\textbf{H}_0$) not only actively contributes to the inertial range energy transfer but also modifies the input energy injection ($\varepsilon$) coming from large scales to the inertial range. In the next section, we numerically investigate if ferrofluid turbulence shows an inertial range energy cascade and eventually calculate the cascade rate.
\vspace{0.5cm}

{\centering \section{\label{sec:level5}Numerical method and simulation details}}

\begin{figure}[H]
\centering
\includegraphics[width=1.0\linewidth]{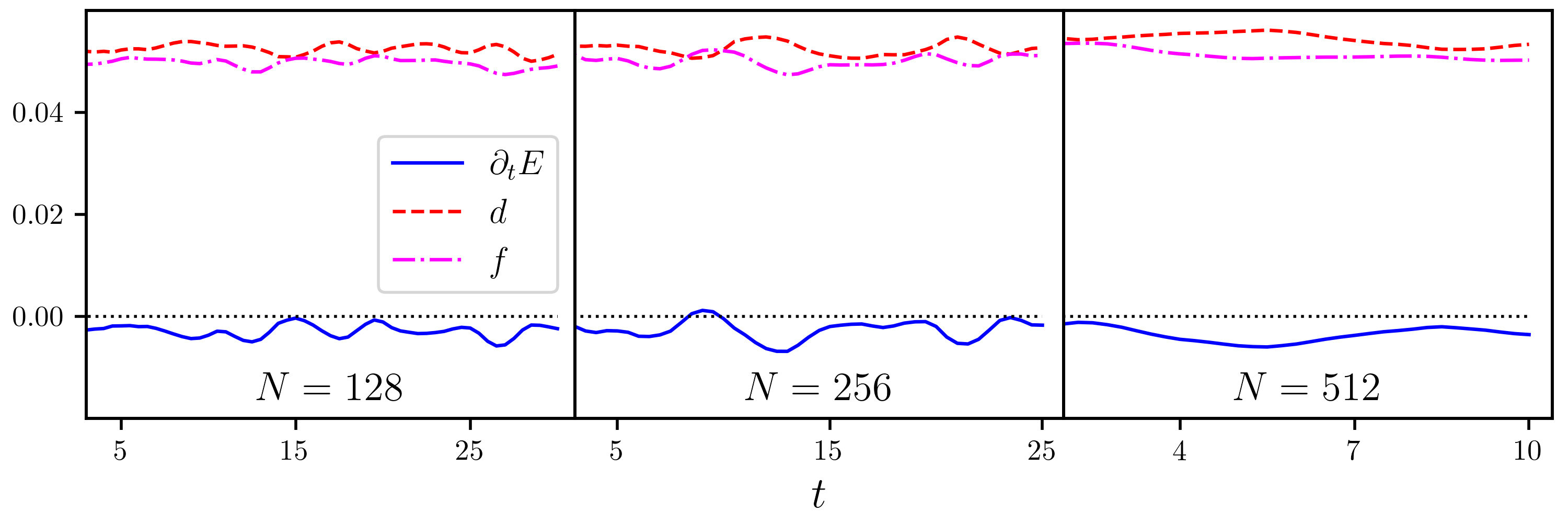}
\caption{{\small Time variation of $ \partial_t E$, energy injection rate $f = \langle\bm{f}\cdot\bm{v}\rangle$ and dissipation rate $d$ for the external field strength $H_0 = 0.1$.
See Run 1a (left), Run 2a (center) and Run 3a (right) from Table \ref{tab} for simulation parameters.}}
\label{fig: 3}
\end{figure}

We perform three-dimensional direct numerical simulations (DNS) of Eqs.~\eqref{22a} and \eqref{22b} using a pseudospectral method with periodic boundary conditions. The box length is taken to be 2$\pi$ with $N$ grid points in each direction. We used resolutions $128^3$, $256^3$, and $512^3$ for the simulations. The aliasing error is removed by a standard $2/3$-dealiasing method, thus limiting the maximum available wavenumber to $N/3$. The code is parallelized using an MPI-based slab decomposition scheme \cite{Mortensen2016}.
The velocity field is initialized from a fully developed pure Navier-Stokes flow. A random initial condition is used for the magnetization field and a uniform time-independent external magnetic field of strength $H_0 = 0.1$ is applied along the $z$ direction. The energy is injected by forcing the momentum evolution equation with a large-scale Taylor-Green forcing $\bm{f}_v \equiv f_0[sin(k_ox)cos(k_oy)cos(k_oz), -cos(k_ox)sin(k_oy)cos(k_oz),0]$, where $k_o = 2$ is the energy-injection scale and $f_0 = 0.5$ is the forcing amplitude. The system is time evolved using a fourth-order Runge-Kutta (RK4) method until a statistical stationary state is achieved.
In the following, using the exact law derived above, we shall numerically investigate if a universal energy cascade of energy is obtained in fully developed ferrofluid turbulence and also the contribution of different nonlinear terms. 
\begin{figure}[H]
    \centering \includegraphics[width=\textwidth]{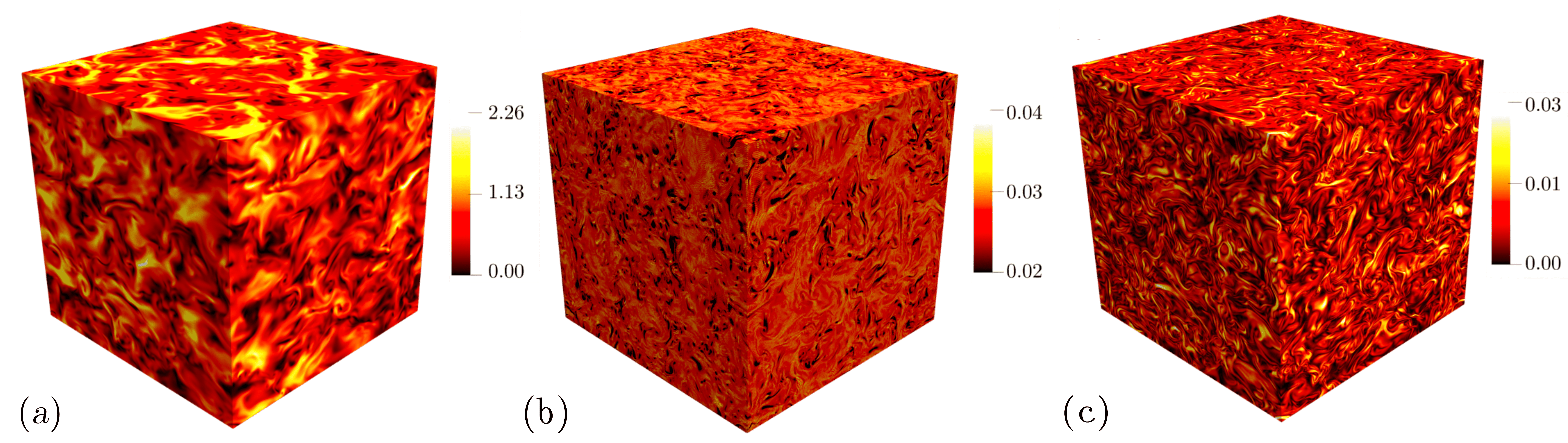}
    \caption{Snapshot of the modulus of (a) velocity ($\bm{v}$)(b) magnetization ($\textbf{M}$) and (c) magnetic field ($\textbf{\~H}$) at simulation time $t = 10$ for Run 3a (Table \ref{tab}).}
         \label{fig:2}
\end{figure}

In Fig. \ref{fig: 3}, we plot the rate of total energy ($\partial_t E = \int \partial_t (E_{kin} + E_{mag}) d \tau$), which is determined by the volume integration over the rate of translational kinetic energy and the rate of work due to magnetization in the presence of an external field of the fluid (see Eq.~\eqref{10}).  As evident from the figure, the average energy dissipation rate ($d$) is balanced by the average energy injection rate ($f$) leading to a statistically stationary state for the total energy i.e., $\partial_t E \approx 0$. All the statistics are done when the statistical steady state is achieved. The presence of structures at all scales in the velocity, magnetization, and magnetic field intelligibly signifies that the turbulence is fully developed (see Fig.~\ref{fig:2}). 
Simulation parameters are summarised in Table \ref{tab}. All the runs are well resolved as the ratio of the maximum wave-number $k_{max} = N/3$ to the Kolmogorov wave number $k_{\eta} = 2 \pi / {\eta}$ where $\eta =  (\nu^3/\varepsilon)^{1/4}$ is greater than 1.  One could expect to better resolve the Kolmogorov scale by increasing the grid size i.e., the ratio $k_{max}/k_{\eta}$ should increase with the increase in grid size. However, at the same time, we are decreasing the fluid viscosity which corresponds to a higher value of $k_{\eta}$. Thus, we could achieve $k_{max}/k_{\eta}\sim 1.5$ for $512^3$ grid points. 

\begin{figure}[H]
\centering
\includegraphics[width=1.0\linewidth]{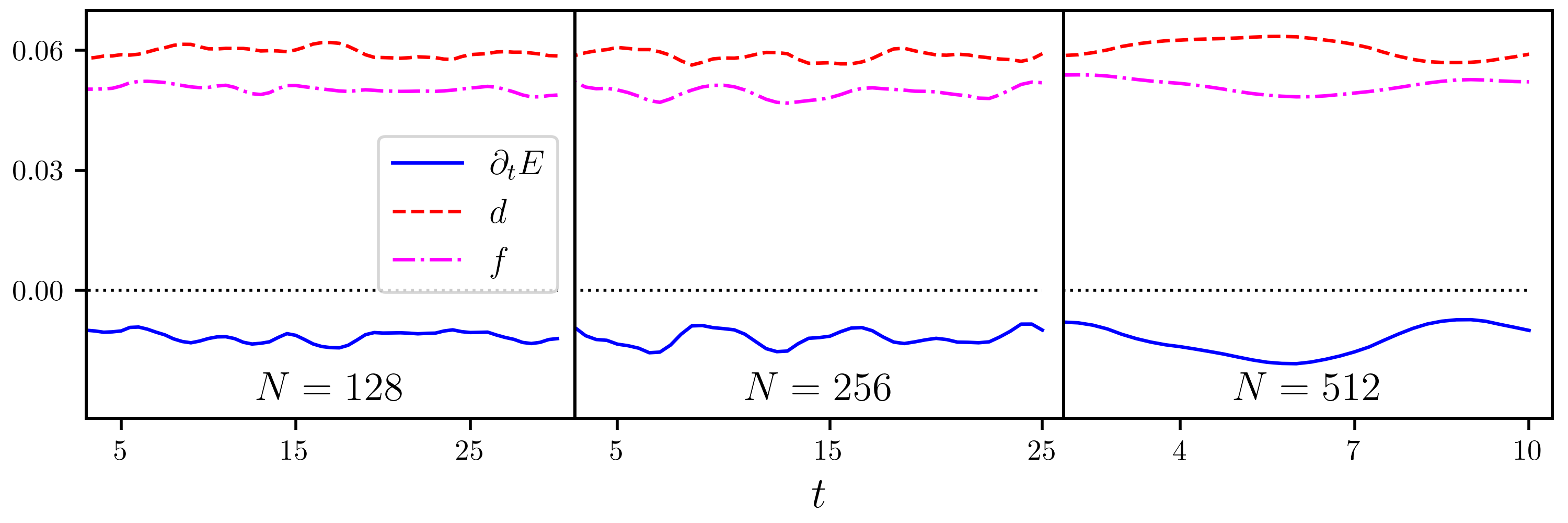} 
\caption{{\small Time variation of $\partial_t E$, $f$ and $d$ for the external field strength $H_0 = 1.0$. See Run 1b (left), Run 2b (center) and Run 3b (right) from Table \ref{tab} for simulation parameters.}}
\label{fig:stat}
\end{figure}

In addition to the aforesaid simulations, we have also performed a series of three simulations for a comparatively strong field strength $H_0 = 1.0$. Rather than searching for a stationary state, we have used the same forcing ($f_0 = 0.5$) as used for the weaker $H_0$ and collected the data for a state where the average energy is not constant but changing at a constant rate in time (see Fig. \ref{fig:stat}). In this case, we investigate whether the kinetic and the total energy still cascade with constant rates. 

\begin{table}[ht]
\caption{\textbf{Simulation parameters}.  $Re$ is the Reynolds number corresponding the box size, $U_{rms}$ is the rms (root mean square) velocity, $\lambda = (5 \int E(k)dk/\int k^2 E(k) dk)^{1/2}$ is the Taylor length scale, $L = (3 \pi/4) {\int E(k)k^{-1}dk}/{\int E(k)dk}$ is the integral length scale, where $E(k)$\, is the energy spectrum. $Re_{\lambda} = U_{rms}\lambda/\nu$ is the Taylor-scale Reynolds number and $Re_L =U_{rms}L/\nu$ is the integral-scale Reynolds number.}
\label{tab}
\begin{tabular}{c c c c c c c c c c c c}
\hline\hline
  \hspace{0.05cm} Run \hspace{0.05cm} & \hspace{0.25cm} $N$ \hspace{0.25cm} & \hspace{0.45cm} $Re$ \hspace{0.45cm} & \hspace{0.15cm} $H_{0}$ \hspace{0.15cm} & \hspace{0.15cm} $U_{rms}$ \hspace{0.15cm} & \hspace{0.4cm} $\Gamma$ \hspace{0.4cm} & \hspace{0.4cm} $\lambda$ \hspace{0.4cm} & \hspace{0.4cm} $L$ \hspace{0.4cm} & \hspace{0.4cm} $\eta$ \hspace{0.4cm} & \quad $Re_{\lambda}$ \quad & \quad $Re_L$ \quad & \hspace{0.05cm} $k_{max}/k_{\eta}$ \hspace{0.05cm} \\
\hline
1a & $128$  & 250& 0.1 & 0.739 & 0.1 & 0.343 & 0.613 & 0.21 & 63 & 113 & 1.432 \\

1b & $128$  & 250& 1.0 & 0.731 & 0.1 & 0.377 & 0.651  & 0.222 & 69 & 118 & 1.508 \\

2a & $256$  & 500 & 0.1 & 0.772 & 0.1 & 0.247 & 0.541 & 0.125 & 96 & 209 & 1.698 \\

2b & $256$  & 500 & 1.0 & 0.776 & 0.1 & 0.284 & 0.57 & 0.133 & 110 & 221 & 1.812 \\

3a & $512$  & 1428 & 0.1 & 0.797 & 0.1 & 0.15 & 0.506 & 0.057 & 172 & 576 & 1.553 \\

3b & $512$  & 1428 & 1.0 & 0.803 & 0.1 & 0.17 & 0.514 & 0.061 & 200 & 589 & 1.662 
\\
\hline \hline
\end{tabular}
\end{table}

For a fully developed turbulent flow, inside the inertial range, a flat region is expected for $ A(\bm{\ell})$ when plotted as a function of increment vector $\bm{\ell}$. The flatness indicates the scale-independent nature of the energy cascade rate and this constant flux rate is equal to the average injection rate i.e., $ A(\bm{\ell}) = const = 2\varepsilon$. 
The average $\langle (\cdot) \rangle$ is calculated over the possible pairs of $\bm{x}$ and $\bm{x}'=\bm{x}+\bm{\ell}$ i.e., $ A(\bm{\ell}) = \langle A (\bm{x},\bm{\ell}) \rangle_{\bm{x}}.$
This average is sufficient for the calculation of all flux terms. However, in practice, exact scaling laws are written assuming statistical isotropy.
In order to achieve that, we vary the increment vector over 73 directions spanned by the base vectors $ \in$ \{($1,0,0$), ($1,1,0$), ($1,1,1$), ($2,1,0$), ($2,1,1$) ($2,2,1$), ($3,1,0$), ($3,1,1$)\} (in units of the grid resolution).
\begin{figure}[H]
  \centering  \includegraphics[width=0.6\textwidth]{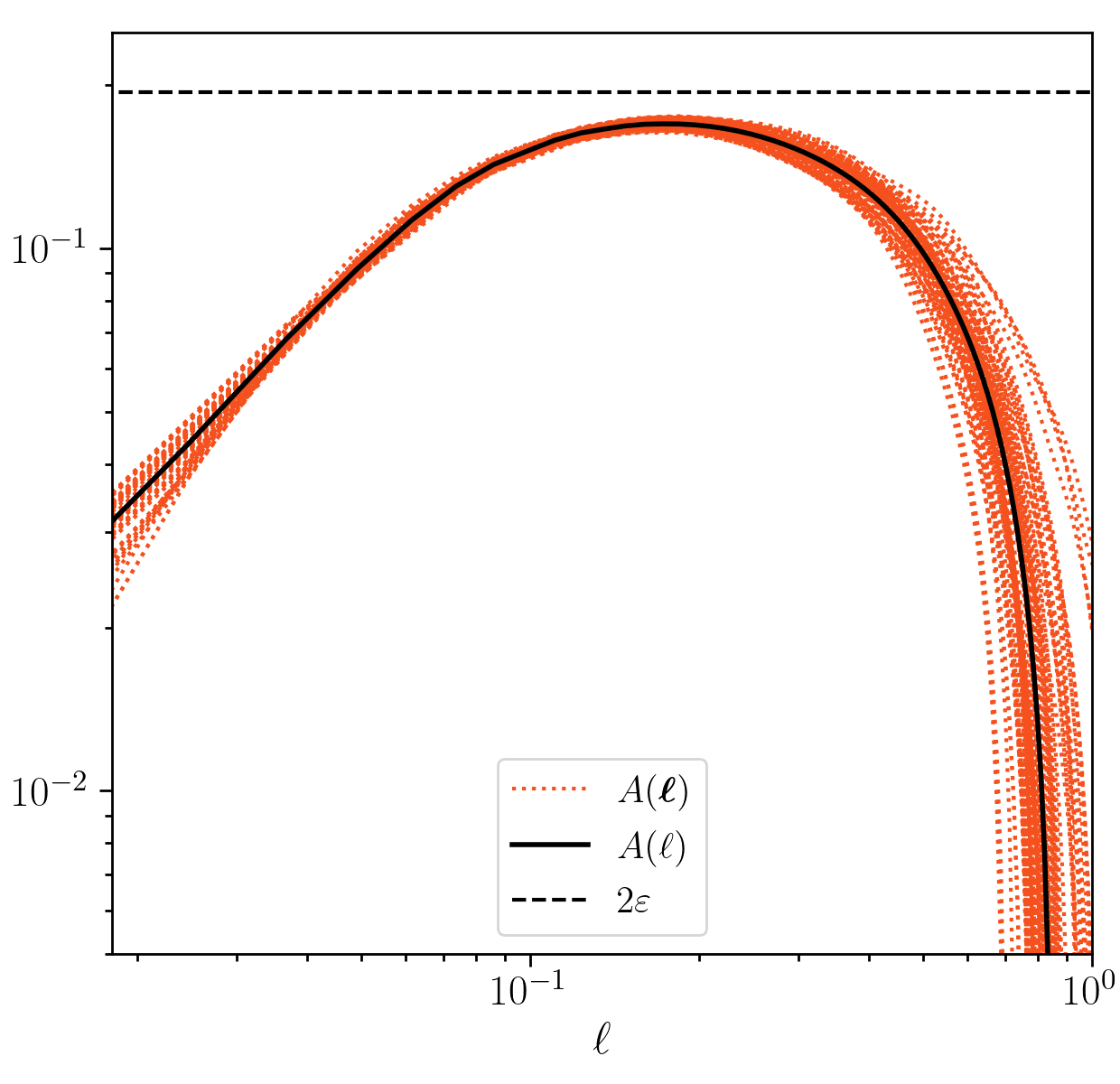}
\caption{Energy cascade rate along all the 73 directions (dotted curves) and the average over all the directions (solid black curve) for a typical simulation (Run 3a).}
\label{all_direcs}
\end{figure}
Finally, a one-dimensional interpolation is done before the final averaging over all the 73 directions, which gives the isotropic $\ A (\ell) = \sum A (\bm{\ell})/73$ \cite{Taylor2003, Nahuel2019, Ferrandthesis}.
Interestingly, in our derived exact relation, $A(\bm{\ell})$ is found to be more or less identical in all 73 directions and hence $A(\ell) \simeq A(\bm{\ell})$ (see Fig.~\ref{all_direcs}). This is because our derived $A(\bm{\ell})$ is according to the alternative form (\cite{Banerjee2018, Banerjee2016, Banerjee2016b}) and is free of any global divergence. 

{\centering \section{\label{sec:level}Results}}

\begin{figure}[H]
  \centering  \includegraphics[width=0.92\textwidth]{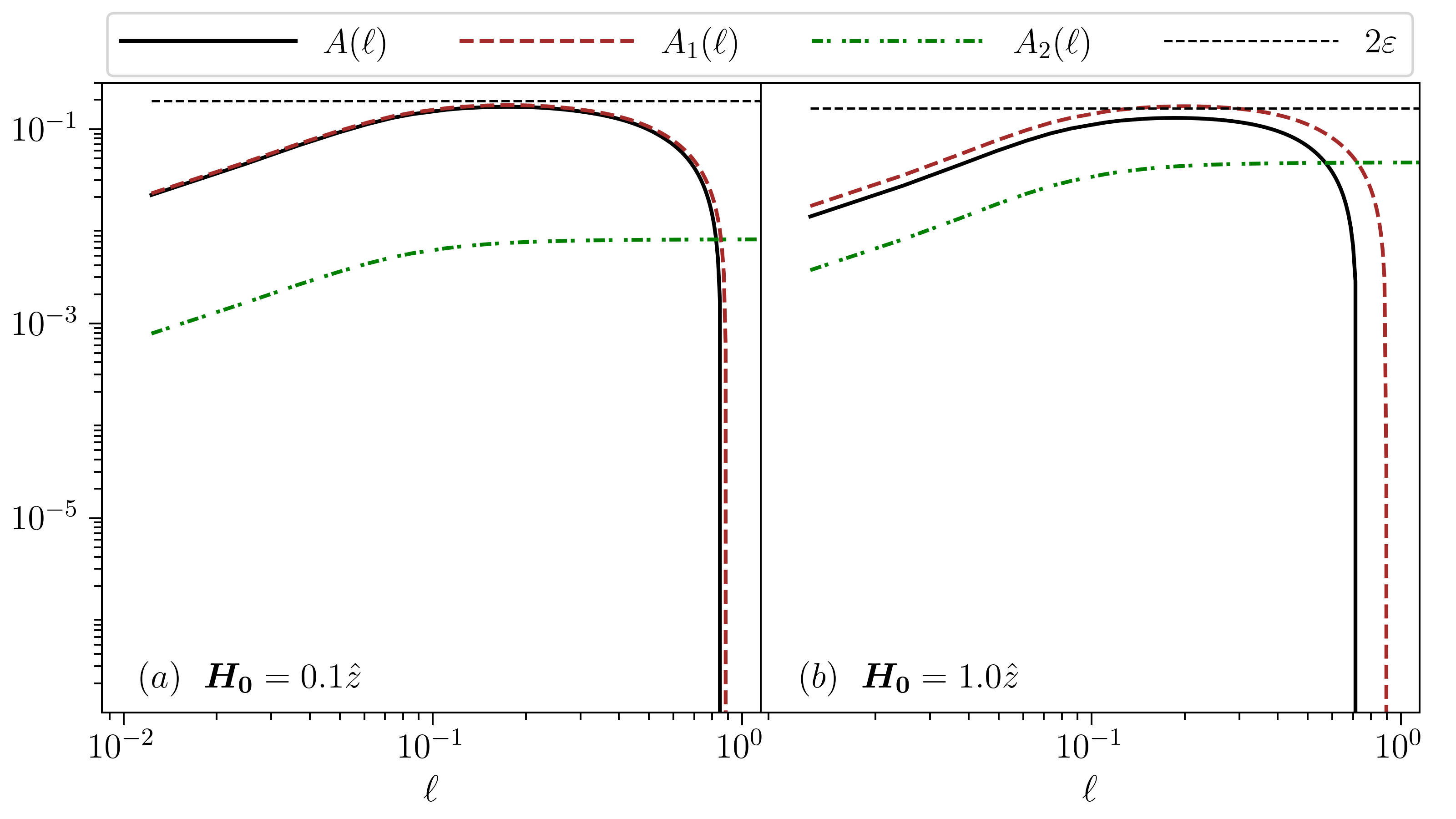}
\caption{Energy cascade rates as a function of $\ell$ for a box with $512^3$ grid points for ${\bf H_0} = 0.1 \hat{z}$ (left) and ${\bf H_0} = 1.0 \hat{z}$ (right).}
\label{flux1}
\end{figure}

In Fig.~\ref{flux1}, we plot $A(\ell)$, $A_1(\ell)$, $A_2(\ell)$ for external field strengths ${H}_0 = 0.1 $ and $ 1.0$. While $A_2(\ell)$ consists of the flux contributions due to the initial external field $\textbf{H}_0$, $A_1(\ell)$ accounts for the contributions due to $\textbf{\~H}$. A flat region in $A(\ell)$ (black solid curve) is clearly found within the intermediate range of scales. It indicates the existence of a scale-independent energy cascade within the internal range and thus numerically validates the Kolmogorov-type of universality for our derived exact relation (Eq.~\eqref{23}) for ferrofluids. 
For statistical stationary state, in addition, the constant scale-independent cascade rate should be equal to twice the average energy injection rate ($\varepsilon$) at large scales. This equality expressed in Eq.~\eqref{23} holds reasonably well for $H_0(= 0.1)$ where we observe $A(\ell)\approx 2\varepsilon$ within the inertial range (see Fig.~\ref{flux1} (a)). This is well expected as we achieved a statistical stationary state for one point average energy with $H_0 = 0.1$. Also, $A_2(\ell)$ is small compared to $A_1(\ell)$ ($\sim 20$ times smaller), and consequently the total energy transfer is mainly governed by $A_1(\ell)$ as $A_1(\ell)\approx A(\ell) \approx 2\varepsilon$. Similar to Fig.~\ref{flux1} (a), both $A(\ell)$ and $A_1(\ell)$ in Fig.~\ref{flux1} (b) become constant for a range of scales thus exhibiting a clear signature of both kinetic and the total energy cascades. 
\begin{figure}[H]
  \centering  \includegraphics[width=0.92\textwidth]{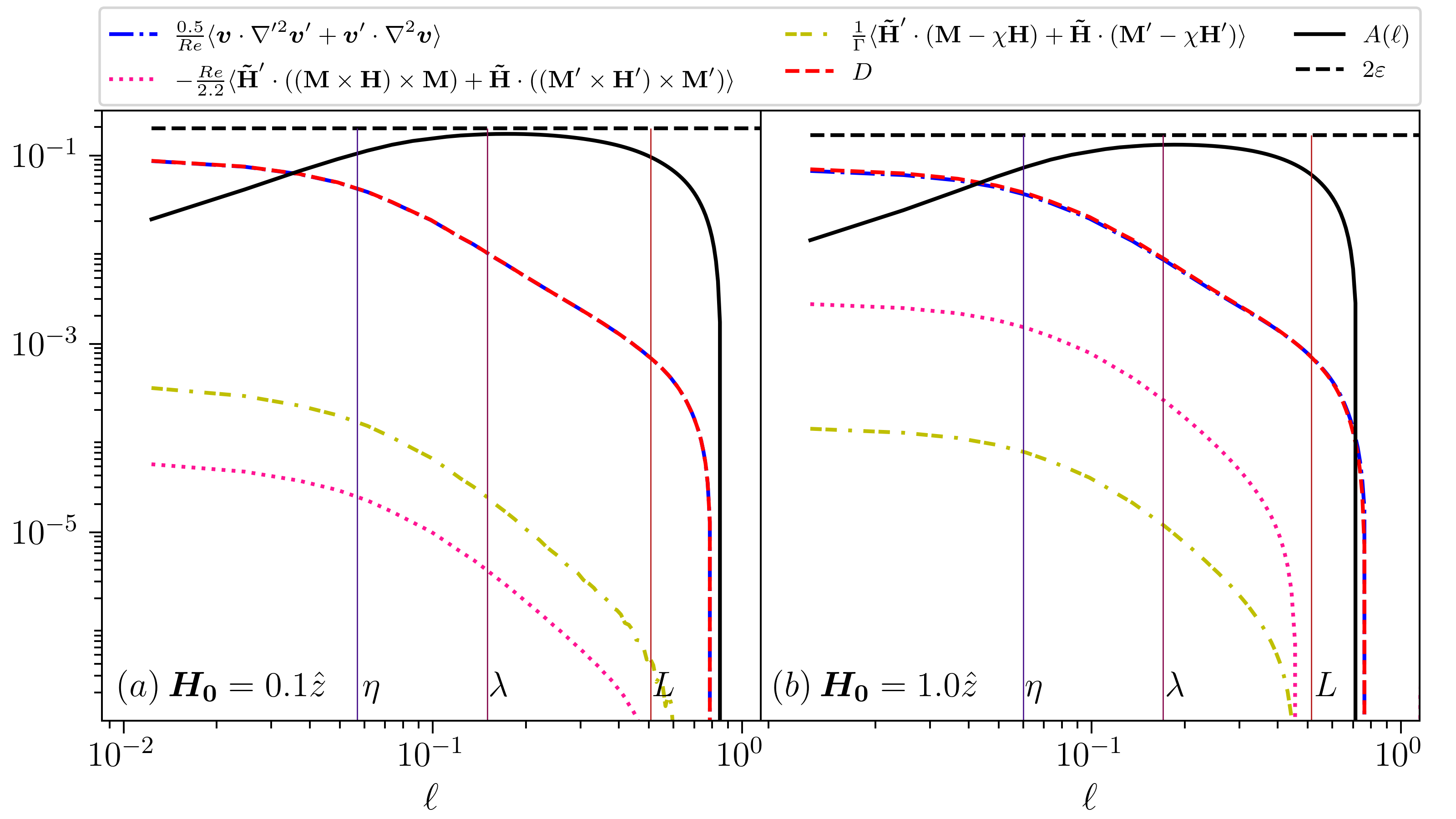}
\caption{Various components of two point energy dissipation rate and $A(\ell)$ as a function of $\ell$ for a box with $512^3$ grid points for ${\bf H_0} = 0.1 \hat{z}$ (left) and ${\bf H_0} = 1.0 \hat{z}$ (right).}
\label{diss}
\end{figure}

\begin{figure}[H]
\centering
\includegraphics[width=1.0\textwidth]
{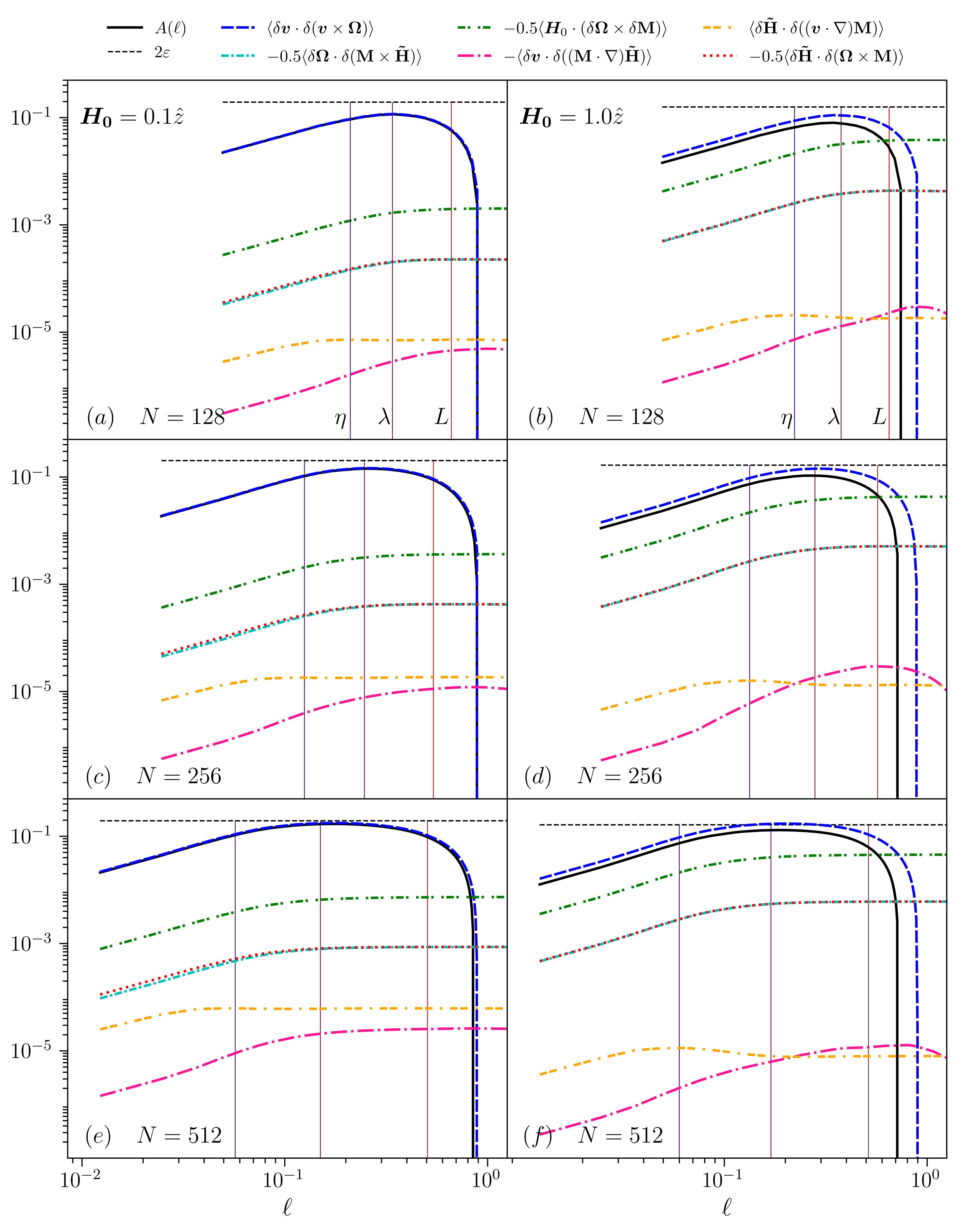}
\caption{Comparison of the energy cascade rates $A(\ell)$ and its components for grid points $128^3$ to $512^3$ (top to bottom) with external magnetic field strength $H_0 = 0.1$ and $H_0 = 1.0$ (left and right). See Table~\ref{tab} for simulation details.}
\label{fig: 4}
\end{figure}
However, for stronger external field ($H_0=1.0$), $A_2(\ell)$ increases and becomes only $\sim 4$ times smaller than $A_1(\ell)$ which remains unaffected by any change in ${H_0}$. Due to negative contribution of $A_2(\ell)$ in $A(\ell)$, the latter decreases with respect to $A_1(\ell)$. Note that, $2 \varepsilon$ also decreases as $H_0$ increases and is slightly dominated by $A_1(\ell)$ in the inertial range as is evident from \ref{flux1} (b). Globally, one can easily see that the exact relation $A(\ell) = 2\varepsilon$ is not closely verified for $H_0 = 1.0$ as there appears a gap between $A(\ell)$ and $2\varepsilon$. This could be due to the fact that scale dependent dissipative terms are no longer negligible in the inertial length scales. In order to probe into this, we have also plotted different contributions from the total dissipation in Fig. \ref{diss}. For both values of $H_0$, all the dissipation effects are found to remain small in the inertial range and become important at small scales. It is therefore natural to understand that the gap between $A(\ell)$ and $2\varepsilon$ is mainly because a proper stationary state was not obtained for $H_0 =1.0$ and the exact law is calculated in the presence of a non zero but constant $\partial_t{R}$. This is particularly interesting as it gives an evidence of energy cascade even in a non-stationary regime where the more general exact relation $A(\ell) = \partial_t{R} + 2\varepsilon$ is satisfied and affirms a constant $A(\ell)$ when $\partial_t{R}$ is a non-zero constant.
In Fig.~\ref{fig: 4}, we plot $ A(\ell) $ along with its components by varying the number of grid points from $128^3$ to $512^3$. As is expected, both for $H_0 = 0.1$ and $H_0 = 1.0$, the extent of the flat region increases with the number of grid points and a better convergence towards $ A(\ell) =2\varepsilon$ is obtained (see Fig.~\ref{fig: 4}~($e$) and Fig.~\ref{fig: 4}~($f$)). 

The extent of the inertial range can be described in terms of the largest energy-containing integral scale ($L$) and the Kolmogorov scale ($\eta$) corresponding to the smallest eddies. As one goes on increasing the resolution (or grid points), the large scale Reynolds number $Re$ increases and so is the ratio ($L/\eta$) thus corresponding to a wider inertial range. From the plots, the increase in the inertial range with grid size is evident from the considerable shift of $\eta$ towards smaller scales whereas $L$ shifts very slightly towards the left. Similar to $\eta$, the Taylor microscale ($\lambda$), which denotes the length scale where dissipation begins to impact turbulent eddies, also appears to migrate towards smaller scales leading to a wider inertial range. All these three length scales are denoted by vertical lines for all the runs plotted in Fig.~\ref{fig: 4}. 
For all the runs, the pure kinetic term  
$\left< \delta \bm{v} \cdot \delta (\bm{v} \times \boldsymbol{\Omega})\right>$ is found to be the most dominating term in $A(\ell)$ (see Table.~\ref{Tab2} for the relative order of magnitude of various terms).  Among other terms, the energy transfer due to $\left< \delta \boldsymbol{\Omega} \cdot \delta (\textbf{M} \times \textbf{\~H}) \right>$ and $\left< \delta \textbf{\~H} \cdot \delta (\boldsymbol{\Omega} \times \textbf{M}) \right>$ increases in magnitude with $H_0$. However, these two contributions cancel each other within the inertial range and thus do not affect the total energy transfer. 
The remaining terms $\langle\delta \textbf{\~H} \cdot \delta((\bm{v} \cdot \boldsymbol{\nabla}) \textbf{M})\rangle$ and $\langle\delta \bm{v} \cdot \delta((\textbf{M} \cdot \boldsymbol{\nabla}) \textbf{\~H})\rangle$ are several orders of magnitude less than the kinetic term and are almost unaffected by the strength of external field. Finally, as mentioned above, the effect of the external field enters through $\langle\textbf{H}_0 \cdot (\delta \boldsymbol{\Omega} \times \delta \textbf{M})\rangle$ which increases with ${H}_0$. 
\begin{table}[ht]
\caption{\small{Order of magnitude of different terms of $A(\ell)$ for grid points $512^3$}}\label{Tab2}
\label{order}
\begin{tabular}{c c c c c c c}
\hline\hline
   \scriptsize{$H_0$} & \scriptsize{$\left< \delta \bm{v} \cdot \delta(\bm{v} \times \boldsymbol{\Omega})\right>$} & \scriptsize{$\left< \delta \bm{v} \cdot \delta((\textbf{M} \cdot \boldsymbol{\nabla}) \textbf{\~H})\right>$} & \scriptsize{$\left<\delta \textbf{\~H} \cdot \delta((\bm{v} \cdot \boldsymbol{\nabla}) \textbf{M})\right>$} & \scriptsize{$\left< \delta \textbf{\~H} \cdot \delta (\boldsymbol{\Omega} \times \textbf{M}) \right>$} & \scriptsize{$\left< \delta \boldsymbol{\Omega} \cdot \delta (\textbf{M} \times \textbf{\~H}) \right>$} & \scriptsize{$\left<\textbf{H}_0 \cdot (\delta \boldsymbol{\Omega} \times \delta \textbf{M})\right>$} \\
\hline
$0.1$ & $0.1$ & $10^{-5}$ & $10^{-5}$ & $10^{-3}$ & $10^{-3}$ & $10^{-2}$
\\
$1.0$ & $0.1$ & $10^{-5}$ & $10^{-4}$ & $10^{-2}$ & $10^{-2}$ & $10^{-1}$ \\
\hline \hline
\end{tabular}
\end{table}

{\centering \section{\label{sec:level6}Discussion and Conclusions}}

\begin{figure}[H]
     \centering  \includegraphics[width=0.85\textwidth]{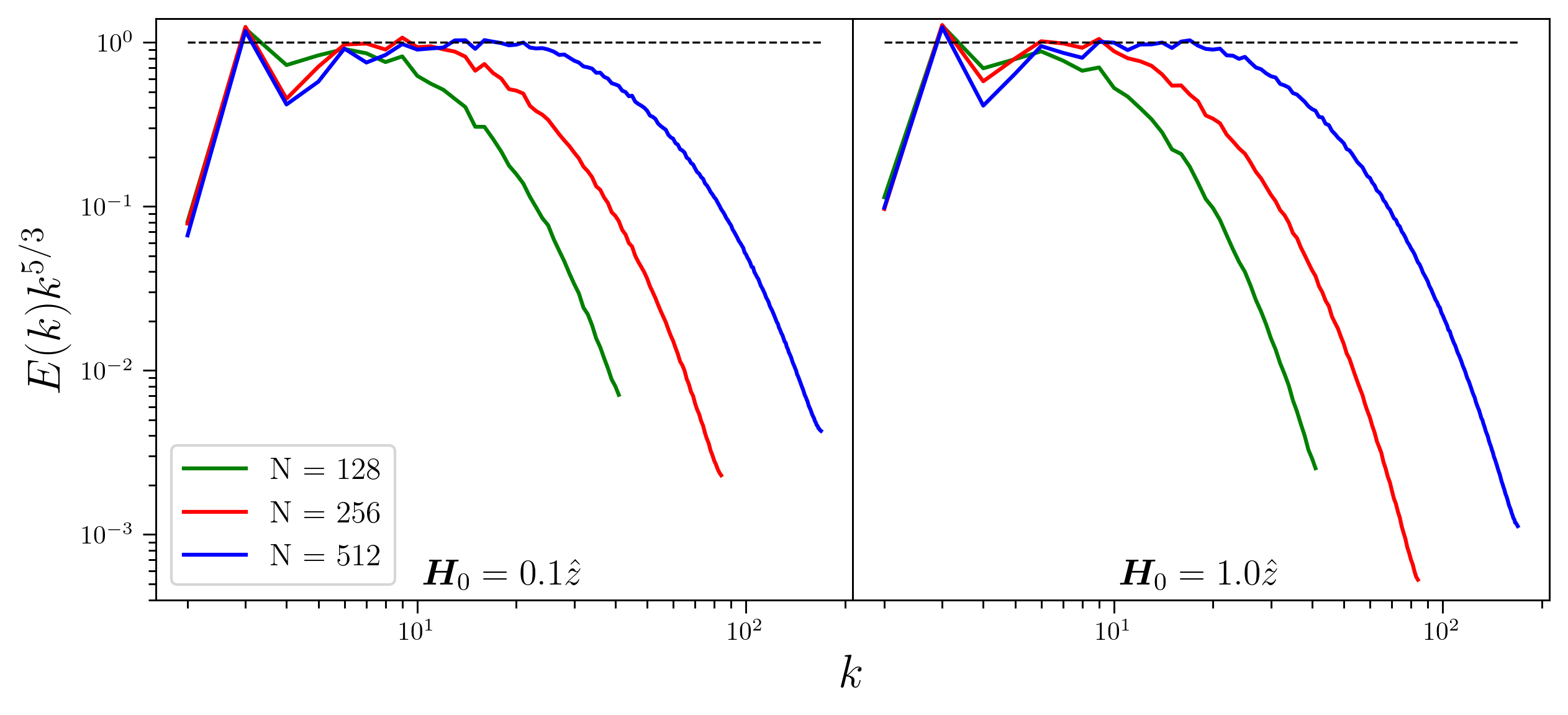}
         \caption{Compensated kinetic energy spectrum for all Runs in Table \ref{tab}.}
         \label{fig:Spectrum}
\end{figure} 
Using two-point statistics, we have derived the exact relation corresponding to the transfer of total energy for three-dimensional incompressible homogeneous ferrofluid turbulence. The exact relation is equally applicable to isotropic and anisotropic flows. For anisotropic flows, unlike the divergence form of the exact relation, the energy cascade rate can be calculated in a straightforward manner without worrying about the geometry of the system. Using direct numerical simulations, we have numerically calculated the cascade rate owing to our derived exact relation. For weak external magnetic field, we have studied the turbulent dynamics of a statistical stationary state where the kinetic energy almost equals the total energy and exhibits inertial range cascade.  For stronger external magnetic field, we have studied the dynamics of a non-stationary regime with a constant energy dissipation rate. Note that, for a strong $H_0$, achieving a statistical stationary state is found to be practically difficult as ferromagnetic particles tend to align themselves quickly along a strong external field thereby leading to a constant leakage of energy from the fluid to the external field. In this case too, both the kinetic and the total energy exhibit constant transfer rate in the inertial range of scales. However, the kinetic energy cascade rate is found to be greater than the total energy cascade rate. 
With the increase in resolution from $128^3$ to $512^3$, we also observed that the flat region between Kolmogorov scale and integral scale increases and a better convergence towards the exact law is achieved. In spectral space, the kinetic energy power spectrum gives a $k^{-5/3}$ spectrum for both values of $H_0$ thereby justifying the existence of a Kolmogorov type cascade for kinetic energy in ferrofluid turbulence (see Fig. \ref{fig:Spectrum}). Similar to the physical space, wider range for $k^{-5/3}$ spectrum is also found for simulations with increasing grid points. Finding power spectra for total energy is, however, not evident for ferrofluid turbulence where the energy density function cannot be written due to the inexact differential form of the magnetic energy part.   

 By the help of the derived exact relation, one can also study the turbulent relaxation in Ferrofluids using the recently proposed Principle of Vanishing Nonlinear Transfer 
 (PVNLT) \cite{Banerjee2023PVNLT}. This principle has recently been used to successfully predict the relaxed states in binary fluid turbulence \cite{pan2024universal} and is currently being implemented to find the relaxation in ferrofluids in a separate study. Complementary to the current study, one can also investigate if the energy conservation is also satisfied in triads and search for the corresponding mode to mode transfers \cite{Banerjee2023fundamental}. Using the said concept, a natural continuation would be to study contributions from local and non-local triads in the energy cascade in ferrofluids in the presence of an external magnetic field of various strengths. 
 The current study can also be extended to the compressible ferrofluid system including the temperature evolution equation into account \citep{Mouraya2023}.
\section*{Data availability}
\vspace{-0.25cm}
All the data presented in the manuscript and the code used in the simulations and analysis are available upon a reasonable request to the corresponding author.
\vspace{-0.5cm}

\section*{Acknowledgments}
The authors thank Arijit Halder and Anando Gopal Chatterjee for useful discussions. The simulation code is developed with the help of parallelization schemes given in Ref. \citep{Mortensen2016}. The simulations are performed using the support and resources provided by PARAM Sanganak under the National Supercomputing Mission, Government of India at the Indian Institute of Technology, Kanpur. SM acknowledges the support for financial assistance of Ph.D. fellowship from CSIR-SRF (Grant No. 09/092(0989)/2018). 

\providecommand{\noopsort}[1]{}\providecommand{\singleletter}[1]{#1}%

\end{document}